\documentstyle[12pt]{article}
\begin{document}
\begin{flushright}
IITM-TH-94-01\\

\vspace{.2cm}
May 1994 ~~~~~~~~
\end{flushright}

\vspace{1cm}

\baselineskip=24pt

\begin{center}
{{\Large \bf Quantum Optimal Control Theory}}\\
\vspace{0.5cm}
{by}\\
\vspace{0.5cm}

{\bf{G.H.Gadiyar\\
Department of Mathematics, Indian Institute of Technology,\\
 Madras 600 036, INDIA.}}\\
\end{center}

\vspace{2.5cm}
\noindent{\bf Abstract}.
The possibility of control of phenomena at microscopic
level compatible with quantum mechanics and quantum field theory is
outlined. The theory could be used in nanotechnology.

\newpage

\noindent {\bf 1.Motivation.}

In technology physical processes are usually
controlled by human intervention. The issue of finding the best in some
sense or optimal control is an issue of great interest in a whole
range of problems. These problems are also of interest to
mathematicians. The classic solution to this problem was given  by
Pontryagin [1]. There have been many extensions and generalizations
but for the extension of the theory to quantum phenomena the work
of Pontryagin turns out to be the best. He has reduced the problem to
one in the calculus of variations and completely solved the problem:
the method is called the maximum principle.

     The issue which will be addressed here is: How does one control
quantum phenomena? At the level of quantum mechanics one has to
respect the uncertainty principle and so the usual classical theory
cannot be naively extended. At the level of quantum field theory
one has to further respect the possibility of second quantization and
creation of particles. It turns out that the difficulties can be
side stepped and the quantum problems are almost as simple as the
classical ones. Mathematical rigor is avoided: the pre-Weierstrass
view that a physically sensible model possesses a mathematically
sensible solution is taken. Rigorous proofs will be provided later.

     The idea is to control microscopic phenomena. For example, can
one minimize the time taken for a certain physical process? The hope
is that in quantum computers with quantum switches, this
theory can be applied to optimize the time to switch from 0 to 1.
It can also be used to invert the population in a laser in an optimal
way. Such control is possible  within the framework of the following
theory.

\vspace{0.5cm}
\noindent {\bf 2.The classical Pontryagin optimal control theory.}

    Consider any controlled process described by a system of ordinary
differential equations
$$
\frac{dx^i}{dt}~=~f^i(x^1,...,x^n;u^1,...,u^r) \ ,\quad i=1\, {\rm  to} \,  n \
. \eqno(1)
$$
Here $x^i$ are coordinates of the process (typically phase space
coordinates.) and $u^r$ are control parameters (usually external forces).
For the process (1) to be defined
$$
  u^j~=~u^j(t) \ , \quad j~=~1\,  {\rm to} \,  r \eqno(2)
$$
are to be specified.
Given
$$
x^i(t_0)~=~x^i_0 \ , \quad i~=~1 \,  {\rm to } \,  n \ ,\eqno(3)
$$
the solution to (1) is uniquely specified.
Usually a functional
$$
J~=~\int^{t_1}_{t_0}f^0(x^1,..,x^n;u^1,...,u^r)dt  \eqno(4)
$$
is to be optimized where $f^0(x^n,u^r)$ is specified.
If $f^0(x^n,u^r)~=~1$, $J$ will be a time optimal problem.

The problem is to optimize (usually minimize) $J$ by tuning the
controls $u^r$ such that the equations (1) are obeyed. Further in
technical applications the $u^r$ are usually constrained by the fact
that the `forces' cannot be arbitrarily large. Constraints like
$$|u^r|~\leq~C^r$$
are typical.

An example will fix the ideas:
Consider a harmonic oscillator subject to a force. Bring it to rest
in least time. Here the problem would be framed as
\begin{eqnarray*}
\frac{dx^1}{dt} &=& x^2\\
\frac{dx^2}{dt} &=& -x^1~+~u \ , \quad |u|~\leq~1
\end{eqnarray*}
Here $x^1~=~x$ is the position, $x^2~=~p~=~{\displaystyle {\frac{dx}{dt}}}$ is
the momentum and $u$ is the force.
The functional $J~=~{\displaystyle {\int^{t_1}_{t_0} \, 1 \,dt}}$ is to be
minimized.

     The solution can be got by the Pontryagin principle but
the physics of the solution can be seen as follows: The force $u$
should be applied always in the opposite direction to the velocity
and should be of maximum magnitude. When the velocity changes sign the force
jumps to the opposite direction again opposing the velocity. Thus
the control jumps is piecewise continuous as a function of time.
This is `bang-bang' control as it is popularly called.

The theory of Pontryagin is summarized in the following results.
Consider in addition to
$$\frac{dx^i}{dt}~=~f^i(x,u) \ ,  \quad  i~=~1\,  {\rm to} \, n \ ,$$
the equation
$$\frac{dx^0}{dt}~=~f^0(x,u)$$
coming from $J$,
that is, now consider
$$\frac{dx^i}{dt}~=~f^i(x,u) \quad ,  \quad i~=~0 \, {\rm to} \, n \quad ({\rm
{\underline {not \, 1 \, {\rm to }\, n}}}) \ .$$
Take an auxiliary set of variables $\psi_0$ to $\psi_n$
$$\frac{d\psi_i}{dt}~=~-\sum_{\alpha~=~0}^{n}\frac {\partial f^\alpha}
{\partial x^i} \, \psi_\alpha \quad ,  \quad i~=~0~{\rm to}~n.$$
This system is linear and homogeneous and we can combine the
equations in to a Hamiltonian
\begin{eqnarray*}
{\cal H}(\psi,x,u) & = & \sum \psi _\alpha f^\alpha (x,u) \ ,\\
\frac{dx^i}{dt} & = & \frac{\partial {\cal H}}{\partial \psi_i} \ ,
\quad \, \, i~=~0~{\rm to} ~n \ ,\\
\frac{d\psi_i}{dt} & = & -\frac{\partial {\cal H}}{\partial x^i} \ ,
\quad i~=~0~{\rm to} ~n \ .
\end{eqnarray*}
Denote by ${\cal M}(\psi,x)~=~{\displaystyle{{\rm sup}_{u\in U}}} {\cal
H}(\psi,x,u)$
where we take the strict upper bound of $\cal H$ as a function
of $u$ for given $\psi,x$.

\vspace{0.5cm}

\noindent {\bf Theorem.}\, Let $u(t), t_0\leq t\leq t_1$ be a permissible
control.
A necessary condition for $u(t)$ and $x(t)$ to be optimal is that there
is a $\psi (t)$ corresponding to $x(t)$ so that

\noindent (1) given an $t_0~\leq~t~\leq~t_1$
$${\cal H}\left (\psi(t),x(t),u(t)\right )~=~{\cal M}\left (\psi (t),x(t)\right
) \ , \eqno(A)$$
that is $\cal H$ attains a maximum at $u(t)$.

\noindent (2) At the final instant $t_1$,
$$\psi _0(t_1)~\leq~0 \quad , \quad {\cal M}(\psi (t_1),x(t_1))~=~0 \ .
\eqno(B)$$
    Further if $\psi (t),x(t)$ satisfy the equation of motion and
$|u(t)|~\leq~1, \psi _0$ and ${\cal M}(\psi (t),x(t))$ are constants
and the equation (B) can be verified for all $t$, $t_0~\leq~t~\leq~1$ and
not only at $t_1$.

\vspace{0.5cm}

\noindent {\bf Example.}
$$
\begin{array}{lclcl}
{\cal H} &=&  \psi _1 x^2 ~-~  \psi _2 x^1 ~ + ~ \psi _2 u \ ,\\
\\
\displaystyle {\frac{dx^1}{dt}} & = & \displaystyle {x^2 \ , \quad
\frac{dx^2}{dt}}  =  \displaystyle {-x^1~+~u} \ ,\\
\\
\displaystyle {\frac{d\psi _i}{dt}} & = & \displaystyle {\psi _2 \ ,\quad
\frac{d\psi _2}{dt}}  =  \displaystyle {-\psi _1} \ .
\end{array}
$$
\noindent So $\psi _1~=~A\, {\rm sin} (t-\alpha _0)$, $A~>0$ and $\alpha _0$
constant.
${\rm Max}\ {\cal H}(\psi,x,u)~=~{\rm sign} \, \psi _2~=~
\newline {\rm sign} (A {\rm sin} (t-\alpha  _0)).$
Hence the control is `bang-bang' as  intuitively argued earlier.

\vspace{0.5cm}
\noindent {\bf 3. Quantum mechanical controls.}

     We cannot obviously generalize by replacing $x$ and $p$ by $
\hat {x}$ and $\hat {p}$ or even $<\hat {x}>$ and $<\hat {p}>$ as
a little thought will indicate. The way the problem is addressed is to
consider the states $|\phi >$ of the Schr\"{o}dinger equation as
the essential feature. What is meant is that the problem is now
how does one optimally move a system from the initial state $\phi _I$ to
the final state $\phi _F$
subject to some controls $u_i$ and functional $J~=~{\displaystyle {\int^t_{t_0}
f^0(\phi ,u)dt}}$.

     Mathematically the problem is written as:
\begin{eqnarray*}
i\hbar \frac{\partial \phi}{\partial t} & = & -\frac{\hbar ^2}{2m} \frac
{\partial ^2 \phi}{\partial x^2}~+~V(x)\phi ~+~u(x,t)\phi \\
& = & {\cal H}_0 \phi~+~u(x,t)\phi \ .
\end{eqnarray*}
Here $u(x,t)$ is the controlling potential. The problem is to
optimize some functional
$$J~=~\int^{t_1}_{t_0}f^0(\phi ,u)dt \ .$$
The answer to this problem now turns out to be very simple: it can
be recast into a form where the Pontryagin principle can be applied.
This is done as follows.

Consider $|n>$ as eigen functions of ${\cal H}_0 $,
$${\cal H}_0|n>~=~E_n|n> \ .$$
Now $$
i\hbar \frac {\partial \phi}{\partial t}~=~{\cal H}\phi ~=~({\cal H}
{}~+~u(x,t))\, \phi$$ can be rewritten with $\phi (x,t)~=~\sum_{n=o}
^\infty C_n(t)|n>$ as
$$i\hbar \sum_{n=0}^\infty \dot {C}_n(t)|n>~=~{\cal H}_0\sum_{n=0}
^\infty C_n(t)|n>~+~u(x,t)\sum_{n=0}^\infty C_n(t)|n>.$$
Now taking the inner product with $<m|$ \ ,
%$$
\begin{eqnarray*}
i\hbar \sum_{n=0}^{\infty} \dot {C}_n(t) <m|n> &=& E_n C_n(t) <m|n>
{}~+~C_n(t)<m|u|n>\\
i\hbar \dot {C}_n(t) &=& C_n(t)~+~\sum_{m=0}^\infty u_{nm}(t)C_m(t) \ .
\end{eqnarray*}
%$$
This is now a linear system of ordinary differential equations and hence
the Pontryagin theory can be applied! The technical problem is that
the index $n$ runs from zero to infinity: that is a problem for
mathematical analysts to tackle. There is a simple case where we have a
spin half particle: that is simpler and is a nice exercise to do.

\vspace{0.5cm}
\noindent {\bf 4.Quantum field theory.}

The same trick as the earlier section can be applied again!
The procedure is as follows.
$$ \begin{array}{lcl}
\left[\phi (x),\phi(y)\right] & = &  0 \ ,\\
\\
\left[\pi (x),\pi (y)\right] & = & 0 \ ,\\
\\
\left[\phi (x),\pi (y)\right] & = & i \, \delta (x-y) \end{array}
$$

\noindent are the canonical commutation relations.
We can use a ``Schr\"{o}dinger picture"
$$i\frac {\partial}{\partial t}\psi (\phi ,t)~=~{\cal H}(-i\frac
{\delta}{\delta \phi},\phi) \ \psi (\phi ,t)$$
where $\psi$ is a functional.
    Thus one can hope to extend the description of Section 3 to this
situation as well. However two problems exist and have to be
understood carefully. How to deal with Fermions and how to address
renormalization effects. However in many body theory in condensed matter
the method can be pushed through.

\vspace{0.5cm}
\noindent{\bf 5.Conclusion.}

     The control problem and the Pontryagin principle can be easily
extended to the quantum domain. Mathematically the problem is
similar though questions of analysis (infinity and convergence)
have to be addressed more carefully.

\newpage
     Technologically it is tempting to think of possible applications
to

\noindent (1) laser population inversion

\noindent (2) quantum switching in quantum computers.

\vspace{0.5cm}
\noindent{\bf Reference.}

\noindent [1] {\it The mathematical theory of optimal processes},
       Pontryagin, Boltyanskii Gamkrelidze and Mishchenko.

\end{document}